\nonstopmode

\newcommand{\HF}{(HF)$_{\infty}$}
\newcommand{\HCl}{(HCl)$_{\infty}$}
\newcommand{\DCl}{(DCl)$_{\infty}$}
\newcommand{\monHCl}{HCl}
\newcommand{\eV}{\U{eV}}
\newcommand{\angstrom}{\U{\hbox{\AA}}}
\newcommand{\degree}{{}^{\circ}}
\newcommand{\ie}{i.e.{}}
\newcommand{\eg}{e.g.{}}
\newcommand{\etal}{\textit{et al.}}
\newcommand{\U}[1]{\,{\rm{#1}}}
\newcommand{\imag}{{\rm i}}
\newcommand{\mat}[1]{\hbox{\boldmath{$#1$}\unboldmath}}
\newcommand{\Sum}{\sum\limits}
\newcommand{\Lim}{\lim\limits}
\newcommand{\unitmatrix}{\mat{\mathbbm{1}}}
\newcommand{\differential}{\>\textrm{d}}
\newcommand{\bra}[1]{\left<\right.\!#1\!\left.\right|}
\newcommand{\ket}[1]{\left|\right.\!#1\!\left.\right>}

\documentclass[prb,showpacs,twocolumn,nopreprintnumbers]{revtex4}
\usepackage{graphicx,subeqnarray,bbm,bm}
\usepackage[nativepdf,bookmarks,bookmarksopen,bookmarksnumbered,raiselinks,%
pdftitle={Quasiparticle band structure of infinite hydrogen fluoride
and hydrogen chloride chains},%
pdfauthor={Christian Buth},%
pdfsubject={Solid state physics},%
pdfkeywords={ab initio, Green's function, ADC, CO-ADC, crystal
orbital, algebraic diagrammatic construction, Wannier function,
quasiparticle band structure, hydrogen fluoride, hydrogen chloride,
crystal, infinite chain}]{hyperref}

\begin{document}
\title{Quasiparticle band structure of infinite hydrogen fluoride
and hydrogen chloride chains}
\date{October 18, 2006}
\author{Christian Buth}
\altaffiliation[Present address: ]{Argonne National Laboratory,
9700~South Cass Avenue, Argonne, Illinois~60439, USA}
\email[electronic mail: ]{Christian.Buth@web.de}
\affiliation{Max-Planck-Institut f\"ur Physik komplexer Systeme,
N\"othnitzer Stra\ss{}e~38, 01187~Dresden, Germany}

\begin{abstract}
We study the quasiparticle band structure of isolated,
infinite~\HF{} and \HCl{} bent (zigzag) chains and
examine the effect of the crystal field on the energy levels
of the constituent monomers.
The chains are one of the simplest but realistic models of
the corresponding three-dimensional crystalline solids.
To describe the isolated monomers and the chains, we set out
from the Hartree-Fock approximation,
harnessing the advanced Green's function methods
\emph{local molecular orbital algebraic diagrammatic
construction}~(ADC) scheme and \emph{local crystal
orbital ADC}~(CO-ADC) in a strict second order approximation,
ADC(2,2) and CO-ADC(2,2), respectively, to account
for electron correlations.
The configuration space of the periodic correlation
calculations is found to converge rapidly only requiring
nearest-neighbor contributions to be regarded.
Although electron correlations cause a pronounced
shift of the quasiparticle band structure of the chains
with respect to the Hartree-Fock result,
the bandwidth essentially remains unaltered
in contrast to, \eg, covalently bound compounds.
\end{abstract}

%
%
%

\preprint{arXiv:cond-mat/0606081}
\maketitle

\section{Introduction}

Hydrogen fluoride~\cite{Atoji:CS-54,Habuda:NM-71,%
Otto:BE-86,Panas:HF-93,Berski:DHF-98}
and hydrogen chloride~\cite{Sandor:CS-67,Sandor:CC-67}
are representatives of hydrogen-bonded crystals.
This type of bonding~\cite{Hamilton:HB-68,%
Pauling:NC-93,Scheiner:HB-97,Hadzi:TT-97,Karpfen:CE-02}
is weak and essentially preserves the
electronic structure of the constituent HF and
HCl~monomers, respectively.
Hydrogen fluoride forms strong hydrogen bonds, whereas
hydrogen chloride forms only weak hydrogen bonds.
Therefore, the two compounds are good candidates to study
the quasiparticle band structure of hydrogen-bonded
periodic systems.
The structures of both crystals are very similar
at low temperatures;
the monomers are arranged in terms of
weakly interacting parallel zigzag chains.
The isolated infinite chain is considered to be a simple
but realistic one-dimensional model for the three-dimensional
solids.~\cite{Blumen:CC-76,Blumen:EB-77,Blumen:EBS-77,%
Liegener:AI-87,Berski:PHF-97,Berski:PHF-98,Buth:BS-04,%
Buth:MC-05,Buth:HB-06}

The isolated \HF{} and \HCl{}~chains have been studied
carefully;~\cite{Zunger:BS-75,Blumen:CC-76,Blumen:EB-77,Blumen:EBS-77,%
Liegener:AI-87,Berski:PHF-97,Berski:PHF-98,Buth:BS-04,Buth:MC-05,Buth:HB-06}
particularly the ground state has been
examined, \eg, Refs.~\onlinecite{Buth:BS-04,Buth:MC-05,Buth:HB-06}
give a detailed \emph{ab initio} analysis of the
electron correlation contributions and the basis
set dependence of the binding energy.
The Hartree-Fock band structure of~\HF{} is discussed in
Refs.~\onlinecite{Zunger:BS-75,Blumen:CC-76,Blumen:EB-77,%
Liegener:AI-87,Berski:PHF-97}
and the one of the \HCl{} chain is
investigated in Refs.~\onlinecite{Blumen:CC-76,Blumen:EB-77,%
Blumen:EBS-77,Berski:PHF-98}.
Correlation corrections for~\HF{} have been obtained
for the center ($\Gamma$~point) and the edge (X~point)
of the Brillouin zone by Liegener and Ladik~\cite{Liegener:AI-87}
who use
many-particle perturbation theory in the M\o{}ller-Plesset partitioning.%
\footnote{This theory (and its derivates) has been applied
successfully to infinite chains by a number of
authors.~\cite{Suhai:QP-83,Liegener:TO-85,Ladik:QT-88,Ladik:PS-99,%
Hirata:MB-00,Pino:LT-04}
Note that we referred to it as outer valence Green's functions~(OVGFs) in
Refs.~\onlinecite{Buth:MC-05,Buth:AG-05}.
Yet this denomination indicated only the underlying approximations
of~OVGFs, \ie, a perturbation expansion of the self-energy in
conjunction with a diagonal Green's function matrix,
but not the renormalization involved.~\cite{Cederbaum:TA-77,Niessen:CM-84}
Hence, this nomenclature is imprecise and shall be discontinued.}
No correlation corrections have been determined for~\HCl{} so far.

To carry out accurate \emph{ab initio} correlation calculations
for periodic systems, we set out from the \textsc{wannier}
program~\cite{Shukla:EC-96,Shukla:WF-98} which directly
produces Wannier orbitals by solving modified Hartree-Fock
equations.~\cite{Shukla:EC-96,Shukla:WF-98,Buth:MC-05,Buth:AP-up}
To account for electron correlations, an advanced Green's function
method, the crystal orbital algebraic diagrammatic construction~(CO-ADC)
scheme for Wannier orbitals~\cite{Buth:MC-05,Buth:AG-05} is used.
It has been devised recently on the basis of the well-established
ADC scheme for atoms, molecules, and
clusters.~\cite{Schirmer:PP-82,Schirmer:GF-83}
It is a method that approximates the Feynman-Dyson perturbation
series for the self-energy and contains sums of certain
proper and improper diagrams to infinite order.
The scheme sets out from the equations in terms of
Bloch orbitals~\cite{Deleuze:SC-03,Buth:MC-05,Buth:AG-05} and
expresses them in Wannier representation which
allows one to specify efficient cutoff criteria for
the configuration space.
The well-known robustness of the ADC method even facilitates
to study strong electron correlations which typically
occur when inner valence electrons are
treated.~\cite{Cederbaum:CE-86,Buth:IO-03}
Particularly, they have been examined in series of oligomers of increasing
length which model infinite chains.~\cite{Deleuze:SC-92,Deleuze:CE-96,%
Deleuze:SB-96,Deleuze:PB-96,Deleuze:VO-99}
Yet in this work, we focus on the quasiparticle (or main)
states for which a one to one correspondence to Hartree-Fock
one-particle states can be established.
Like the one-particle states, they form a so-called
quasiparticle band structure.~\cite{Fulde:EC-95,Fulde:WF-02}
The CO-ADC theory~\cite{Buth:MC-05,Buth:AG-05} has been
implemented on top of \textsc{wannier} in terms
of the \textsc{co-adc} program.~\cite{co-adc-04}

This article is structured as follows:
The ADC and CO-ADC schemes in local representation are introduced
in Sec.~\ref{sec:theory}, whereas Sec.~\ref{sec:comp} describes
geometries, basis sets, cutoffs, and the computer programs employed.
We devote Sec.~\ref{sec:monomer} to the discussion of
the energy levels of the HF and HCl monomers paying particularly
attention to the impact of electron
correlations and their implications.
The quasiparticle band structure of the
infinite chains is investigated in Sec.~\ref{sec:polymer}.
Conclusions are drawn in Sec.~\ref{sec:conclusion}.

\section{Theory}
\label{sec:theory}

In order to determine the energy levels of atoms, molecules,
clusters, and crystals, we set out from the
Schr\"odinger equation with the full nonrelativistic
Hamiltonian neglecting nuclear motions;
the ground state is obtained in restricted, closed-shell
Hartree-Fock approximation.~\cite{Ashcroft:SSP-76,Pisani:HF-88,Ladik:QT-88,%
Callaway:QT-91,Fulde:EC-95,Pisani:QM-96,Ladik:PS-99}
From the orbital energies, the ionization potentials~(IPs)
and electron affinities~(EAs) follow immediately by
Koopmans' theorem;~\cite{Szabo:MQC-89,%
McWeeny:MQM-92,Helgaker:MES-00}
the energy levels are simply given by their negative value,
\ie, -IP and -EA.

Taking electron correlations into account, the energy levels
now result from the poles of the one-particle
Green's function (or particle propagator)~\cite{Mattuck:FD-76,%
Fetter:MP-71,Schirmer:GF-83,Gross:MP-91,Callaway:QT-91}
in energy representation;
it is expressed, using atomic units, in the spectral or Lehmann
representation by~\cite{Mattuck:FD-76,Fetter:MP-71,Gross:MP-91}
\begin{equation}
  \label{eq:p-spectrum}
  \begin{array}{rcl}
    G_{pq}(\vec k,\omega) &=& \Sum_{n \in \{N+1\}}
    \frac{y_p^{(n)}(\vec k) \; y_q^{(n)*}(\vec k)}
         {\omega + A_n(\vec k) + \imag \eta} \\
    &&{} + \Sum_{n \in \{N-1\}} \nonumber
    \frac{x_p^{(n)}(\vec k) \; x_q^{(n)*}(\vec k)}
         {\omega  + I_n(\vec k) - \imag \eta}
  \end{array}
\end{equation}
and depends on the crystal momentum~$\vec k$.~\cite{Ashcroft:SSP-76,%
Callaway:QT-91,Calais:TS-95,Deleuze:SC-03}
The residue amplitudes are~$y_p^{(n)}(\vec k) =
\bra{\Psi_0^N} \hat c_{\vec k \, p} \ket{\Psi_n^{N+1}(\vec k)}$
and $x_p^{(n)}(\vec k) = \bra{\Psi_n^{N-1}(-\vec k)}
\hat c_{\vec k \, p} \ket{\Psi_0^N}$, where $\hat c_{\vec k \, p}$ destroys the
electron in the Hartree-Fock Bloch orbital~$\psi_{\vec k \, p}(\vec r \, s)$.
Here $\vec r$ and $s$ correspondingly denote spatial and
spin coordinates.
The negative values of the pole positions of~$G_{pq}(\vec k,\omega)$
are the~IPs and the EAs, $I_n(\vec k)$ and $A_n(\vec k)$, respectively.
The infinitesimal~$\eta > 0$ in Eq.~(\ref{eq:p-spectrum})
is needed to render the Fourier transformation
between time and energy variables convergent.

The pole strengths or spectroscopic factors are defined
by~\cite{Cederbaum:TA-77}~$|x_p^{(n)}(\vec k)|^2$ and
$|y_p^{(n)}(\vec k)|^2$ for~IPs and EAs, correspondingly.
They quantify the relative spectral intensities
measured in (inverse) photoelectron experiments
assuming the sudden approximation.~\cite{Cederbaum:TA-77,Cederbaum:CE-86}
Summing over all Bloch orbitals yields the orbital independent
pole strengths which are given by~\cite{Niessen:VI-92}
\begin{equation}
  \label{eq:polestrength}
  \begin{array}{rcl}
    P_+^{(n)} &=& \Sum_{\vec k \, p} |y_p^{(n)}(\vec k)|^2 \\
    P_-^{(n)} &=& \Sum_{\vec k \, p} |x_p^{(n)}(\vec k)|^2 \;.
  \end{array}
\end{equation}
The inequalities $0 \leq |y_p^{(n)}(\vec k)|^2,
|x_p^{(n)}(\vec k)|^2, P_{\pm}^{(n)} \leq 1$ are satisfied.
As~$P_{\pm}^{(n)} = 1$ holds for the $n$th IP and EA, respectively,
obtained in Hartree-Fock approximation via Koopmans
theorem,~\cite{Szabo:MQC-89,McWeeny:MQM-92,Helgaker:MES-00}
formula~(\ref{eq:polestrength}) is a measure for the
one-particle character of the $n$th state.

To evaluate the particle propagator~(\ref{eq:p-spectrum}),
the self-energy~$\Sigma_{rs}(\vec k, \omega)$ with respect to
the residual interaction---here the difference between the
full Hamiltonian and the Fock operator---is introduced
via the Dyson equation~\cite{Fetter:MP-71,Mattuck:FD-76,%
Schirmer:GF-83,Gross:MP-91,Callaway:QT-91}
\begin{equation}
  \label{eq:Dyson}
  \mat G(\vec k, \omega) = \mat G^0(\vec k, \omega) + \mat G^0(\vec k, \omega)
    \, \mat \Sigma(\vec k, \omega) \, \mat G(\vec k, \omega) \; ,
\end{equation}
where~$G^0_{pq}(\vec k, \omega)$ denotes the
Green's function of the noninteracting system.
The self-energy naturally decomposes into an
energy independent part, the static self-energy,
and an energy dependent part, the dynamic
self-energy,~\cite{Cederbaum:TA-77,Schirmer:GF-83}
\begin{equation}
  \mat \Sigma(\vec k,\omega) = \mat \Sigma^{\infty}(\vec k)
  + \mat M(\vec k,\omega) \; ,
\end{equation}
with~$\Lim_{\omega \to \pm \infty} \mat M(\vec k,\omega) = \mat 0$.
The dynamic self-energy also breaks down into two contributions:
\begin{equation}
  \label{eq:Mplusminus}
  \mat M(\vec k, \omega) = \mat M^+(\vec k, \omega)
  + \mat M^-(\vec k, \omega) \; .
\end{equation}
The static and the dynamic self-energies are
related by~\cite{Cederbaum:TA-77,Schirmer:SE-89,Weikert:BL-96}
\begin{equation}
  \label{eq:staticS}
  \begin{array}{rcl}
    \Sigma_{pq}^{\infty}(\vec k) &=& \Sum_{\vec k^{\prime}}
      \Sum_{r,s} V_{\vec k \, p \; \vec k^{\prime} \, r \; [\vec k \, q \;
      \vec k^{\prime} s]} \Bigl [ \frac{1}{2\pi\imag} \oint \Sum_{u,v} G^0_{su}
      (\vec k^{\prime}, \omega) \\
    &&{} \times (\Sigma^{\infty}_{uv}(\vec k^{\prime}, \omega)
      + M_{uv}(\vec k^{\prime}, \omega)) \,
      G_{vr}(\vec k^{\prime}, \omega) \differential \omega \Bigr ] \; .
  \end{array}
\end{equation}
Here $V_{\vec k \, p \; \vec k^{\prime} \, r \; [\vec k \, q \; \vec k^{\prime} s]}$
are antisymmetrized two-electron integrals~\cite{Szabo:MQC-89,%
McWeeny:MQM-92,Helgaker:MES-00}
and the contour integration runs along the real axis and closes
in the upper complex $\omega$~plane.~\cite{Cederbaum:TA-77,%
Schirmer:SE-89,Weikert:BL-96}
Formula~(\ref{eq:staticS}) can be evaluated using the Dyson expansion
method~(DEM).~\cite{Schirmer:SE-89,Buth:MC-05,Buth:AG-05}
Due to the relation~(\ref{eq:staticS}), it is sufficient to
concentrate on the approximation of the dynamic
self-energy because the static part can be obtained
from it afterwards.

The algebraic diagrammatic construction~(ADC) scheme
is used to express the dynamic self-energy in
a nondiagonal representation,
\begin{equation}
  \label{eq:ADC-form}
  M_{rs}^{\pm}(\vec k,\omega)
    = \vec U_r^{\pm\dagger}(\vec k) \> [\omega \, \unitmatrix
      - \mat K^{\pm}(\vec k) - \mat C^{\pm}(\vec k)]^{-1} \>
      \vec U_s^{\pm}(\vec k) \; ,
\end{equation}
the so-called general algebraic form or ADC~form.
The $n$th order approximation of the ADC
is constructed by inserting the perturbation ansatz
in terms of the residual interaction~$\mat U^{\pm}(\vec k)
= \mat U^{\pm\,(1)}(\vec k) + \mat U^{\pm\,(2)}(\vec k) + \cdots$
and $\mat C^{\pm}(\vec k) = \mat C^{\pm\,(1)}(\vec k)
+ \mat C^{\pm\,(2)}(\vec k) + \cdots$ into Eq.~(\ref{eq:ADC-form})
and expanding the resulting formula into a geometric
series.~\cite{Schirmer:GF-83}
Comparing the resulting terms with the terms arising
in the Feynman-Dyson perturbation expansion of the dynamic self-energy,
analytical expressions can be found for the
matrices showing up in Eq.~(\ref{eq:ADC-form}).
Thereby, frequently linear combinations of the expressions
of a few diagrams have to be formed.
The construction works because both series have
the same analytic structure.

To evaluate the equations for the dynamic self-energy,
we switch to a representation in terms of generalized
Wannier orbitals.
Such a local representation is particularly beneficial to treat
crystals with a large unit cell.~\cite{Ladik:QT-88,Fulde:EC-95,%
Ladik:PS-99,Fulde:WF-02}
Suitable schemes to express the crystal momentum dependent
ADC~form~(\ref{eq:ADC-form}) using Wannier orbitals are discussed
in Refs.~\onlinecite{Buth:MC-05,Buth:AG-05}.
In this work, we exclusively use the fully
translational symmetry adapted form in Eq.~(29)
of Ref.~\onlinecite{Buth:AG-05}.

When changing to Wannier orbitals, one has to keep in mind
that the Fock matrix is no longer diagonal.
The off-diagonal matrix elements are most efficiently accounted
for by a diagrammatic expansion in addition to the perturbative
evaluation of the two-electron interaction.~\cite{Buth:MC-05,Buth:AG-05}
The resulting approximation schemes are denoted by CO-ADC($m$,$n$).
This notation indicates that the contributions which
(partly) involve one-electron interactions are treated
in order~$m$ and the contributions that are exclusively
given by two-electron interactions are treated in order~$n$.
In this article, we resort to the lowest order
CO-ADC(2,2) approximation in Eq.~(32)
of Ref.~\onlinecite{Buth:AG-05}.
Then, in Eq.~(\ref{eq:Mplusminus}) of this work,
$M_{rs}^+(\vec k, \omega)$~is associated with
two-particle-one-hole~($2p1h$)~configurations and
$M_{rs}^-(\vec k, \omega)$~involves
two-hole-one-particle~($2h1p$)~configurations.

With the translational symmetry adapted Fock matrix~$\mat F(\vec k)$
and explicit expressions for the static and
the dynamic self-energy, the energy levels of a crystal
can be determined for the crystal momentum~$\vec k$;
they are given by the eigenvalues~$\mat E(\vec k)$ of
the Hermitian band structure matrix,
\begin{widetext}
\renewcommand{\arraystretch}{1.5}%
\begin{equation}
  \mat B(\vec k) = \left(
  \begin{array}{ccc}
    \mat F(\vec k) + \mat\Sigma^{\infty}(\vec k)
    & \mat U^{+\dagger}(\vec k) & \mat U^{-\dagger}(\vec k) \\
    \mat U^+(\vec k) & \mat K^+(\vec k) + \mat C^+(\vec k) & \mat 0 \\
    \mat U^-(\vec k) & \mat 0 & \mat K^-(\vec k) + \mat C^-(\vec k)
  \end{array}
  \right) \; ,
\end{equation}
\renewcommand{\arraystretch}{1}%
\end{widetext}
which are obtained by solving the eigenvalue problem
\begin{equation}
  \label{eq:solvepADC}
  \mat B(\vec k) \mat X(\vec k) = \mat X(\vec k) \mat E(\vec k),
  \qquad \mat X^{\dagger}(\vec k) \, \mat X(\vec k) = \unitmatrix \; .
\end{equation}
Here $\mat X(\vec k)$ denotes the eigenvectors of~$\mat B(\vec k)$.
Its matrix elements, $X_{pn}(\vec k)$, represent the residual
amplitudes~$y_p^{(n)}(\vec k)$ or $x_p^{(n)}(\vec k)$ in
Eq.~(\ref{eq:polestrength}) depending on whether they are
of~$N+1$ or of~$N-1$ type.

As crystals are infinite, the convergence properties of the lattice sums
occurring in the matrix elements of~$\mat B(\vec k)$ have to be examined.
Careful analyses of the sums in expressions which are similar to ours
are carried out in Refs.~\onlinecite{Deleuze:SD-95,Deleuze:SC-95,%
Nooijen:AL-97,Sun:CL-97,Sun:CB-98}.
Although convergent, these sums have to be truncated suitably using a
cutoff procedure because the configuration space, \ie, the number
of~$2p1h$ and $2h1p$~configurations considered to form~$\mat B(\vec k)$,
needs to be restricted.~\cite{Fulde:EC-95,Buth:MC-05,Buth:AG-05}
It can be achieved by realizing that
the importance of individual configurations
for the description of the energy levels of a crystal
can be measured by the magnitude of the one-
and the two-electron integrals involved.
This criterion allows a very fine grained, individual selection
of configurations according to a given cutoff
threshold.
The procedure is termed configuration selection.~\cite{Buth:MC-05,Buth:AG-05}

Having set up the band structure matrix,
its partial diagonalization is carried out efficiently
using a complex-Hermitian band-(or block-)Lanczos
algorithm.~\cite{Buth:MC-05,Buth:AG-05}
The computer code is based on the complex-symmetric version of
Sommerfeld that, in turn, is an adaption of the real-symmetric
implementation by Meyer.~\cite{Sommerfeld:BL-04}
Band Lanczos methods are iterative algorithms that utilize
the product of a matrix, here~$\mat B(\vec k)$,
with a few vectors to transform the original matrix into a
band diagonal form which subsequently can be diagonalized
efficiently.~\cite{Meyer:BL-89,Weikert:BL-96,Sommerfeld:SI-00}
The band-Lanczos algorithm usually converges to
the spectral envelope already after
a few matrix times vector operations.
This corresponds to only moderately sized band diagonal
matrices.~\cite{Meyer:BL-89,Weikert:BL-96}

A quasiparticle point of view is assumed for
the outer valence regime and the lowest virtual
states, \ie, we focus on those states which
are essentially of one-particle character with
only a moderate adjustment due to electron
correlations.~\cite{Fulde:EC-95,Fulde:WF-02}
The number of quasiparticle bands is the same
as the number of Hartree-Fock bands and a clear
association between both can be made.
To identify the quasiparticle states of a crystal
in the spectrum of~$\mat B(\vec k)$, we resort to
the orbital independent pole strength~(\ref{eq:polestrength})
extracting only states with $P_{\pm}^{(n)} > 0.7$.
In analogy to the energies of one-particle states,
the quasiparticle energy levels group into bands with respect to
the crystal momentum, forming a so-called quasiparticle band structure.

A local orbital based ADC scheme for isolated molecules
can easily be devised setting out from the CO-ADC
equations in Wannier representation:~\cite{Buth:MC-05,Buth:AG-05}
let the unit cell of a primitive cubic crystal lattice with
a macroscopic lattice constant be occupied by a
given molecule.
Then only Fock matrix elements and two-electron integrals
within a unit cell are nonnegligible.
Consequently, all states of the crystal are $N_0$-fold
degenerate and the band structure matrix has only to
be diagonalized at the $\Gamma$~point to obtain all
distinct energy levels of the
crystal.~\cite{Buth:MC-05,Buth:AG-05}
These levels can be identified with the energy levels
of the isolated molecule.
By this line of argument, it is justified to drop the
lattice sums and lattice vectors totally in
the CO-ADC equations to obtain analytical expressions
for the local molecular orbital ADC scheme.
The dependence on the crystal momentum quantum number
vanishes and the resulting matrix~$\mat B$ is now termed
energy level matrix.
Similar to the nomenclature for the methods
based on crystal orbital ADC, the corresponding approximative
schemes for molecules are referred to as ADC($m$,$n$).

\section{Computational details}
\label{sec:comp}

\begin{figure}
  \includegraphics[width=\hsize,clip]{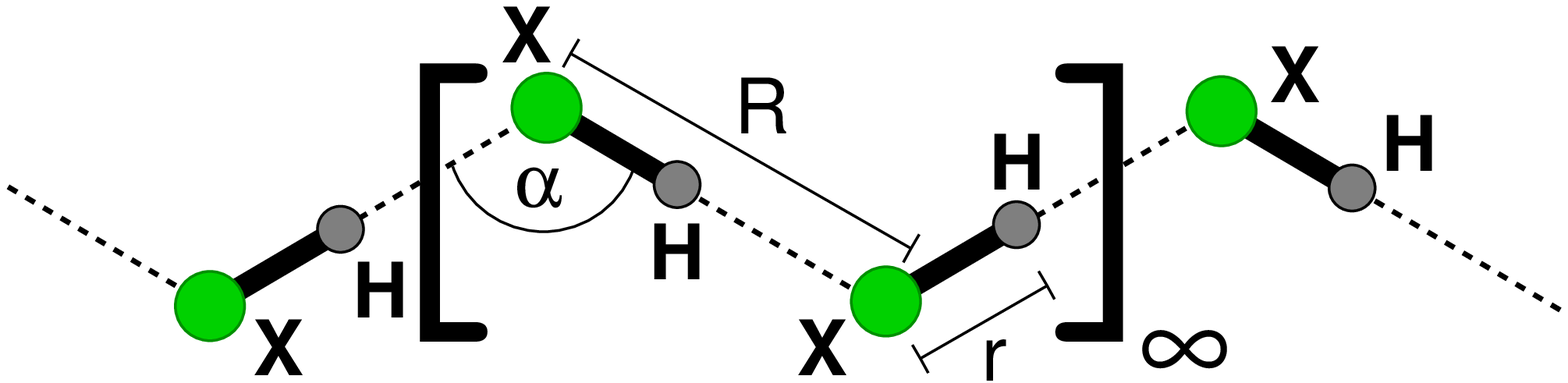}
  \caption{(Color) Structure of infinite~\HF{} and \HCl{}
           zigzag chains with~$X ={}$F, Cl, respectively.}
  \label{fig:polygeom}
\end{figure}

Hydrogen fluoride and hydrogen chloride monomers are
diatomic molecules of~$C_{\infty v}$ symmetry.
The experimental values for the internuclear distance
are~$0.91680 \angstrom$ for HF and $1.27455 \angstrom$
for HCl.~\cite{Huber:MS-79}
Upon crystallization to a three-dimensional solid, the
HF and HCl~monomers arrange in long zigzag chains.
The structure of the single infinite \HF{} and \HCl{}~chain
[Fig.~\ref{fig:polygeom}] is determined by three parameters,
the H---$X$~distance~$r$, the $X \cdots X$~distance~$R$, and the
angle~$\alpha = \angle({\rm H}X{\rm H})$ with X${}={}$F, Cl.
The following experimental values for the parameters are
taken: $r = 0.92 \angstrom$, $R = 2.50 \angstrom$, and
$\alpha = 120 \degree$ for \HF~\cite{Atoji:CS-54,Holleman:IC-01}
and $r = 1.25 \angstrom$, $R = 3.688 \angstrom$, and
$\alpha = 93.3 \degree$ for \DCl{}.~\cite{Sandor:CS-67}
Structural information for HCl crystals is not available.
Yet HCl and DCl crystals have very similar lattice
constants and are considered to be isomorphous.~\cite{Sandor:CS-67}
In this study, we use the cc-pVDZ basis
set~\cite{Dunning:GBS-89,Woon:GBS-93,basislib-04}
to represent~H, F, and Cl~atoms in the compounds.%
\footnote{In Refs.~\onlinecite{Buth:BS-04,Buth:MC-05}, the
cc-pVDZ basis set~\cite{Dunning:GBS-89,Woon:GBS-93,basislib-04}
was found to yield rather inaccurate ground-state binding energies
of \HF{} and \HCl{}~chains;
they deviated in both cases
%
%
%
%
%
%
by~$\approx 0.05 \U{eV}$ from the best value for the
binding energy [Tab.~4 of Ref.~\onlinecite{Buth:BS-04}].
In this study, however, such accuracy of the energy levels
would be very good.
By these numbers, we do not want at all to imply that the
computations presented here are that accurate but
rather indicate the significantly lowered demands
to the quality of the basis set.
The difference between the results from
ADC(2,2) computations and the ADC(3) method~\cite{Schirmer:GF-83}
in Tabs.~\ref{tab:HF_levels} and \ref{tab:HCl_levels}
is typically larger than the basis set error for the ground-state
binding energy.}

A self-consistent solution of modified Hartree-Fock equations is carried
out with the \textsc{wannier} program,~\cite{Shukla:EC-96,Shukla:WF-98}
which directly yields Wannier
orbitals.~\cite{Shukla:EC-96,Shukla:WF-98,Buth:MC-05,Buth:AP-up}
A finite cluster of unit cells is utilized by \textsc{wannier}
serving as a support for the Wannier orbitals in the origin cell.
It consists of the origin cell and up to
fourth nearest-neighbor cells (nine unit cells altogether)
for \HF{} and up to third nearest-neighbor cells (seven unit
cells altogether) for the \HCl{}~chain.
To carry out Hartree-Fock calculations for
molecules, we let each unit cell of a one-dimensional
lattice with a constant of~$1 \U{\mu m}$ comprise a monomer.
This ensures that interactions with neighboring unit cells,
which contain periodic images of the molecule, are negligible.
The Wannier orbitals in the origin cell can thus
be identified with the orbitals of the isolated monomer.

As \textsc{wannier} determines only occupied Wannier
orbitals, translationally related projected atomic orbitals,
so-called crystal projected atomic
orbitals,~\cite{Albrecht:PR-04,Buth:MC-05}
have been devised to be used as virtual Wannier functions.%
\footnote{The intermediate overcompleteness of the
virtual space representation is inherent to the
projected atomic orbital theory.
It is resolved in the \textsc{wannier} program~\cite{Shukla:EC-96,Shukla:WF-98}
following Werner and co-workers~\cite{Hampel:LT-96,Knowles:AI-00}
by removing the projected atomic orbitals which correspond to the smallest
eigenvalues of the overlap matrix formed in
the origin cell.~\cite{Albrecht:PR-04,Buth:MC-05}}
They are based on the projected atomic orbitals
introduced by Pulay~\cite{Pulay:LD-83} and Saeb\o{}
and Pulay~\cite{Saebo:LT-93} and implemented by
Werner and co-workers~\etal.~\cite{Hampel:LT-96,Knowles:AI-00}
The \textsc{wannier} program
produces pseudocanonical Wannier orbitals, \ie,
the occupied block of the Fock matrix in the
origin cell is diagonalized.
Therefore, in the case of molecules occupying a primitive
cubic crystal with a macroscopic lattice constant,
only the virtual block in terms of the projected atomic
orbitals contains off-diagonal matrix elements.

The \textsc{wannier} program~\cite{Shukla:EC-96,Shukla:WF-98}
also performs the transformation
of the Fock matrix elements and the two-electron integrals
from a representation in terms of one-particle basis
functions to the Wannier representation.
These quantities are the only ones which enter
subsequent correlation calculations
by the local molecular orbital ADC and the
local crystal orbital ADC methods.
In this study, we apply the lowest order approximation,
ADC(2,2) and CO-ADC(2,2), to molecules and infinite chains, respectively.
To this end, we remove the explicit spin dependence
of the equations involved, following the arguments of
von Niessen~\etal.~\cite{Niessen:CM-84}
The resulting equations for the two methods are
implemented in terms of the \textsc{co-adc}
program.~\cite{co-adc-04}

Both ADC(2,2) and CO-ADC(2,2) employ an additional
perturbative expansion of the off-diagonal terms
of the Fock matrix in local orbitals
with respect to their canonical orbital based
counterparts.
This introduces an extra approximation
using ADC(2,2) in relation to ADC(2)
whose accuracy has been critically assessed in
Ref.~\onlinecite{Buth:MC-05} for a single HF~molecule.
The outer valence IPs and the smallest EAs are found
to be well described by the correlation method.
Yet it is not able to describe the
strong electron correlations which are observed
in the inner valence region (F$\,2s$~dominated states)
of the HF~monomer in the same way as the
ADC(2)~scheme that, nevertheless, is very inaccurate in this region,
too.~\cite{Buth:MC-05}
This limitation of the method does not imply a constraint
for this work because we focus exclusively on the main
states of the systems.

\begin{figure}
  \centering
  \includegraphics[clip,width=\hsize]{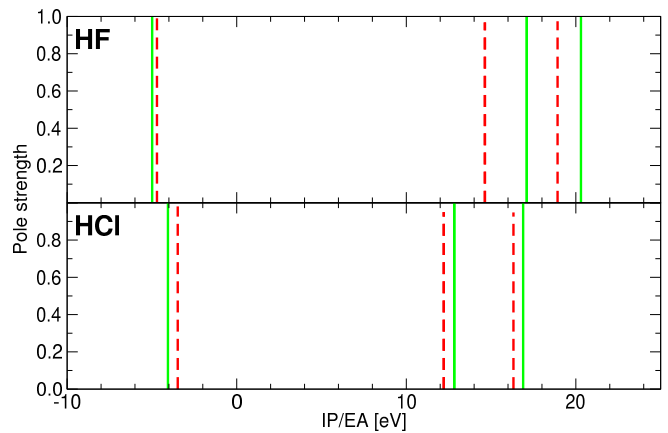}
  \caption{(Color) Ionization potentials and electron
           affinities of a~HF and a HCl molecule where
           IP${} \geq 0$ and EA${} < 0$.
           The solid green lines denote Hartree-Fock values
           (Koopmans' theorem~\cite{Szabo:MQC-89,%
           McWeeny:MQM-92,Helgaker:MES-00}), while
           dashed red lines depict ADC(2,2)~results.}
  \label{fig:HF_HCl_levels}
\end{figure}

\begin{table}[tb]
  \begin{ruledtabular}
  \begin{tabular}{crrr}
    Level No. & $-\varepsilon$ & IP$_{2,2}$/EA$_{2,2}$ & IP$_{3}$/EA$_{3}$ \\
    \hline
    3    &  20.30 &  18.92 &  19.64 \\
    4, 5 &  17.11 &  14.64 &  15.73\medskip \\
    6    &  -4.99 &  -4.70 &  -4.67
  \end{tabular}
  \end{ruledtabular}
  \caption{Ionization potentials and electron affinities of a HF molecule.
           They are given in Hartree-Fock approximation by the
           negative value of the orbital energies~$-\varepsilon$;
           the ADC(2,2) values are denoted by~IP$_{2,2}$/EA$_{2,2}$.
           The reference values~IP$_{3}$/EA$_{3}$ are obtained
           with the ADC(3) scheme.~\cite{Schirmer:GF-83}
           All data are given in electronvolts.}
  \label{tab:HF_levels}
\end{table}

\section{HF and \protect\monHCl{} monomers}
\label{sec:monomer}

Before we turn to the infinite chains in Sec.~\ref{sec:polymer},
we investigate the ionization potentials~(IPs) and the
electron affinities~(EAs) of isolated diatomic~HF and HCl monomers.
This will allow to better interpret the band
structures of the chains later on.
On the abscissa of Fig.~\ref{fig:HF_HCl_levels}, we give the
IPs ($\geq 0$) and EAs ($< 0$) of the molecules while the
ordinate indicates the pole strength~(\ref{eq:polestrength}) of each state.
The data are obtained once directly from the Hartree-Fock
orbital energies by exploiting Koopmans' theorem~\cite{Szabo:MQC-89,%
McWeeny:MQM-92,Helgaker:MES-00}
and the other time in ADC(2,2) approximation.
The numerical values are listed in Tabs.~\ref{tab:HF_levels}
and \ref{tab:HCl_levels} aside from reference IPs and EAs which
were obtained with the ADC(3) scheme~\cite{Schirmer:GF-83}
using \textsc{gamess-uk}~\cite{gamess-uk} for the molecular Hartree-Fock
calculation in conjunction with the ADC~program of
Tarantelli.~\cite{Tarantelli:PC-06}
We do not present data which correspond to the IPs of core states of
F$\,1s$~type and inner valence states of F$\,2s$~type for~HF.
Neither do we show the core states of Cl$\,1s$, Cl$\,2s$, and Cl$\,2p$ types in
conjunction with the inner valence states of~Cl$\,3s$ origin for~HCl.
Above all, only the~EA with the lowest energy in the given basis set
is displayed.
Yet the resonance properties~\cite{Santra:CAP-03}
of the molecules are not investigated here further.

Many of the characteristics of the IPs and the EAs
of the two molecules can already be understood in
the independent particle model.
The outer valence region of the two molecules is formed by
two distinct IPs in the range of~$14$--$22 \U{eV}$ for~HF
and of~$10$--$18 \U{eV}$ for~HCl.
By inspecting the molecular orbitals, we find
the lowest IPs of both monomers to
correspond to an ionization from the two equivalent $\pi$-type lone pairs
on fluorine and chlorine, respectively.
Consequently, they are twofold degenerate.
Ionization from the third $\sigma$-type pair requires more
energy as it is oriented towards the hydrogen atom
and thus is attracted by its positive partial charge.

\begin{table}[tb]
  \begin{ruledtabular}
  \begin{tabular}{crrr}
    Level No. & $-\varepsilon$ & IP$_{2,2}$/EA$_{2,2}$ & IP$_3$/EA$_{3}$ \\
    \hline
    7    &  16.89 &  16.32 & 16.38 \\
    8, 9 &  12.83 &  12.21 & 12.30\medskip \\
    10   &  -4.06 &  -3.48 & -3.44
  \end{tabular}
  \end{ruledtabular}
  \caption{Ionization potentials and electron affinities of a HCl molecule.
           The symbols are chosen as in Tab.~\ref{tab:HF_levels}.
           All data are given in electronvolts.}
  \label{tab:HCl_levels}
\end{table}

In the two spectra, the differences between fluorine and chlorine atoms
basically manifest themselves in two effects.
Firstly, there is an overall shift of all IPs and EAs of HCl to
lower energies with respect to corresponding IPs in~HF because
in chlorine the nuclear charge is shielded additionally
by the lower lying~Cl$\,2s$ and Cl$\,2p$ shells
which are missing in the fluorine atom.%
\footnote{This observation can be compared to studies
of the ionization spectra of the xenon
fluorides where a shift of the inner valence
(and core) IPs to lower energies is observed with increasing number
of fluorine ligands.
There it is traced back to the fact that the addition
of fluorine atoms leads to a reduced screening
of the nuclear charge of the xenon
atom.~\cite{Buth:IO-03}}
Secondly, the larger internuclear distance in
HCl results in a weaker interaction between hydrogen and
chlorine atoms compared with the interaction
between hydrogen and fluorine atoms and causes, in conjunction
with the lower electronegativity of the chlorine atom,
the HCl~monomer to be weaker and more covalently bonded
than the HF~monomer.

As soon as one accounts for electron correlations,
the IPs and EAs of~HF and HCl in Fig.~\ref{fig:HF_HCl_levels}
change considerably;
they shift to lower absolute values with respect
to the corresponding Hartree-Fock values.~\cite{Fulde:EC-95}
The displacement of the IPs in the outer valence region in
ADC(2,2) approximation is noticeably smaller for HCl
compared with~HF;
the contrary holds for the lowest EA.
Inspecting Tabs.~\ref{tab:HF_levels} and \ref{tab:HCl_levels},
one observes that the ADC(2,2) method already provides
in most cases a considerable improvement over the Hartree-Fock
approximation.
However, note that Koopmans' theorem value corresponding to energy
level~3 of the HF molecule is slightly closer to the reference
values than the ADC(2,2) value.
This observation is not very surprising because the HF molecule is well
known to be difficult to describe accurately with quantum
chemical methods.

The fact that the outer valence IPs of~HCl in Hartree-Fock approximation
are closer to the ADC(2,2) values compared with~HF leads to the
conclusion that electron correlations are less important for these
states of~HCl.
The reverse trend holds for the lowest EA of the molecules.
This finding can be ascribed to the fact that the
chlorine atom is notably bigger than the fluorine atom.
Hence, the valence electrons of HF are squeezed to a smaller volume,
leading to a worse performance of Hartree-Fock theory;
it is caused by an insufficient description of the Coulomb hole
around the electrons which is especially important
in the case of large geometrical orbital overlaps.~\cite{Szabo:MQC-89,%
McWeeny:MQM-92,Fulde:EC-95,Helgaker:MES-00}

Electron correlations increase the energy splitting of the IPs
of~HF with respect to the Hartree-Fock result, whereas the splitting
remains nearly the same for~HCl.
Moreover, the approximately uniform shift of the outer valence IPs
of~HF and HCl indicates that
electron correlations for these states are dominated
by the fluorine and the chlorine atom, respectively, and
are essentially unaltered upon formation of the molecular bond.

\section{HF and \protect\monHCl{} chains}
\label{sec:polymer}

\begin{figure}
  \centering
  \includegraphics[clip,width=\hsize]{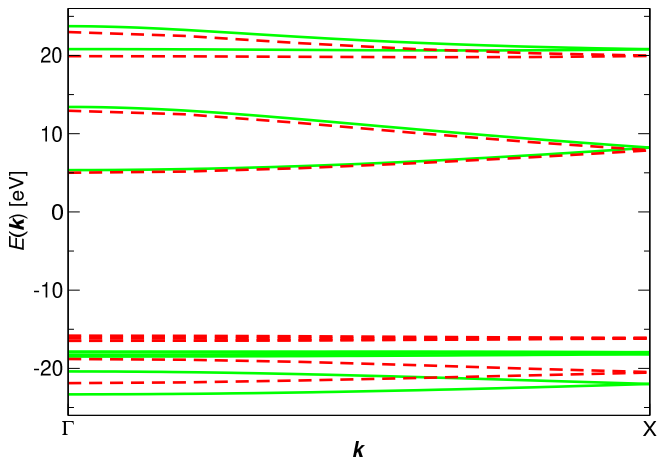}
  \caption{(Color) Band structure of a \HF{} chain.
           Hartree-Fock bands are given by the solid green lines.
           The dashed red lines depict CO-ADC(2,2) quasiparticle bands which
           are determined by accounting
           for $2p1h$ and $2h1p$~configurations that involve
           Wannier orbitals in the origin cell and the nearest- and the
           next nearest-neighbor cells.}
  \label{fig:HF_5cells}
\end{figure}

Having understood the nature of the IPs and EAs of the HF
and HCl~molecules in the previous Sec.~\ref{sec:monomer},
their changes upon crystallization can now be
studied.

\begin{figure}
  \centering
  \includegraphics[clip,width=\hsize]{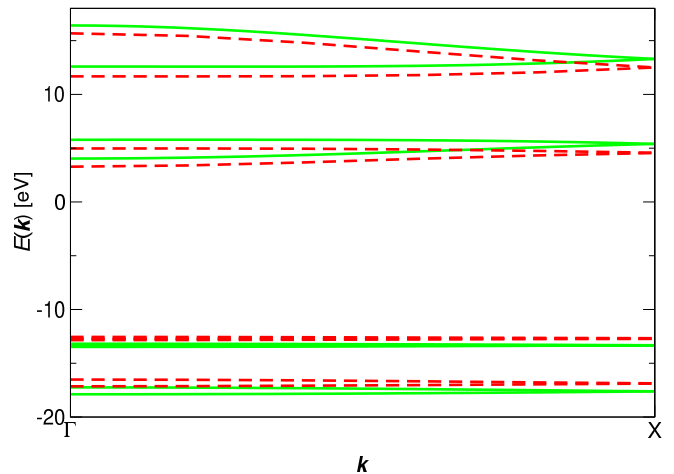}
  \caption{(Color) Band structure of a \HCl{} chain.
           The symbols are chosen as in Fig.~\ref{fig:HF_5cells}.}
  \label{fig:HCl_5cells}
\end{figure}

\begin{figure}
  \centering
  \includegraphics[clip,width=\hsize]{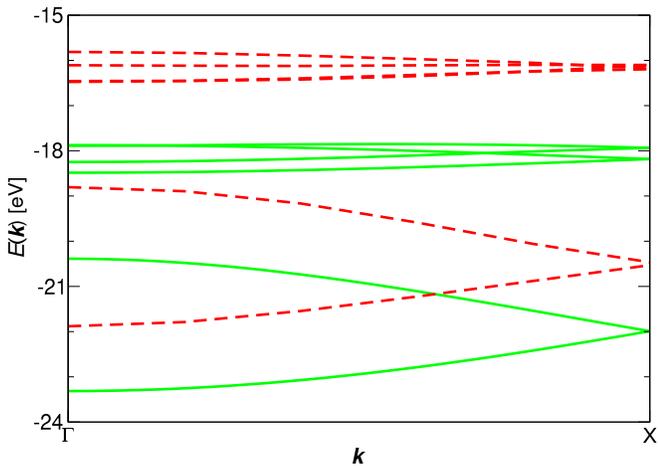}
  \caption{(Color) Valence band structure of a \HF{}~chain.
           Zoom into Fig.~\ref{fig:HF_5cells}.}
  \label{fig:HF_5cells_valence}
\end{figure}

\begin{figure}
  \centering
  \includegraphics[clip,width=\hsize]{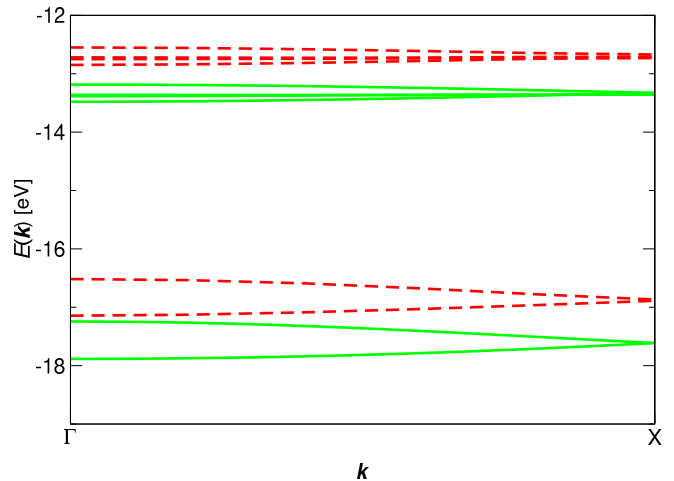}
  \caption{(Color) Valence band structure of a \HCl{}~chain.
           Zoom into Fig.~\ref{fig:HCl_5cells}.}
  \label{fig:HCl_5cells_valence}
\end{figure}

The Hartree-Fock band structure of the \HF{} and the
\HCl{}~chain is displayed in Figs.~\ref{fig:HF_5cells}
and \ref{fig:HCl_5cells}, respectively;
a close-up of the valence bands is shown in
Figs.~\ref{fig:HF_5cells_valence}
and \ref{fig:HCl_5cells_valence}.
Numerical data of the energy levels of both compounds
at the $\Gamma$ and the X~point are given in
Tabs.~\ref{tab:HF_band_XG} and \ref{tab:HCl_band_XG}.%
\footnote{As all calculations with \textsc{wannier}~\cite{Shukla:EC-96,Shukla:WF-98}
and \textsc{co-adc}~\cite{co-adc-04} are performed
without enforcing the rod group symmetry of the chains
explicitly in the equations, a tiny artificial splitting of the energy
levels occurs at the X~point due to a minute breaking of
the symmetry.
We remove it by computing the arithmetic mean between pairs of should-be
degenerate energy levels in Tabs.~\ref{tab:HF_band_XG}
and \ref{tab:HCl_band_XG}.}
As a unit cell of the chains comprises
two monomers, the number of Hartree-Fock bands
is twice the number of Hartree-Fock IPs and EAs
of the monomer [Fig.~\ref{fig:HF_HCl_levels} and
Tabs.~\ref{tab:HF_levels} and \ref{tab:HCl_levels}].
For \HF{}, the four low lying occupied bands, which mainly correspond
to F$\,1s$~core states and F$\,2s$~inner valence states,
are left out.
Similarly, we do not show the energy bands originating from
Cl$\,1s$, Cl$\,2s$, and Cl$\,2p$ core states and the
inner valence of~Cl$\,3s$~character.
Furthermore, only the four lower conduction bands in the cc-pVDZ
basis set~\cite{Dunning:GBS-89,Woon:GBS-93,basislib-04} are
given for both compounds.
Again an examination of the resonance properties~\cite{Santra:CAP-03}
is not pursued further.

The Hartree-Fock bands which are situated energetically around
$-18 \eV$ for~\HF{} and around $-12 \eV$ for~\HCl{} are formed
by the two equivalent $\pi$-type lone pairs of
mostly F$\,2p$ and Cl$\,3p$~character of the isolated
molecules at~$-17.11 \U{eV}$ and $-12.83 \U{eV}$, respectively
[Fig.~\ref{fig:HF_HCl_levels} and Tabs.~\ref{tab:HF_levels}
and \ref{tab:HCl_levels}].
Note that -IP or -EA, correspondingly, gives the energy level.
By inspection of Fig.~\ref{fig:HF_5cells}, we see that the degeneracy
of the molecular orbitals is lifted only slightly
due to the crystal field upon formation
of the infinite chain.
The lowest valence bands cross in Figs.~\ref{fig:HF_5cells_valence}
and \ref{fig:HCl_5cells_valence},
which is allowed by rod group symmetry.

The energy bands of~\HF{} in the range of~$-24 \eV $ to $-20 \eV$
and of~\HCl{} between~$-18 \eV $ and $-17 \eV$ are constituted
by the third outer valence orbital of the isolated monomers
which is also predominantly of F$\,2p$ and Cl$\,3p$~character
with an orbital energies of~$-20.30 \U{eV}$ and of~$-16.89 \U{eV}$,
respectively [Fig.~\ref{fig:HF_HCl_levels} and Tabs.~\ref{tab:HF_levels}
and \ref{tab:HCl_levels}].
The dispersion of these two energy bands is much larger than the
dispersion of the four bands discussed in the previous paragraph.
This indicates that hydrogen bonding in the chains
is mainly mediated by the orbitals of the monomers which
lead to these two outer valence bands.

The Hartree-Fock band structure of the \HF{}~chain
is reported by Berski and Latajka~\cite{Berski:PHF-97}
for a series of basis sets.
Among these, the 6-31+G(d,p)~basis is of most comparable
quality to the cc-pVDZ basis~\cite{Dunning:GBS-89,Woon:GBS-93,basislib-04}
employed in this study;
the plot of the band structure of~\HF{}
in Fig.~3 of Ref.~\onlinecite{Berski:PHF-97}
agrees very well with the plot in
Fig.~\ref{fig:HF_5cells_valence} of this article.
We read off the energy of the top of the valence bands
at the $\Gamma$~point from Fig.~3 of Berski and Latajka;~\cite{Berski:PHF-97}
it lies
%
%
at~$-18.3 \U{eV}$ which is very close to our value
of~$-17.9 \U{eV}$ from Tab.~\ref{tab:cabHF}.
The Hartree-Fock band structure of the \HF{}~chain has also
been studied by Liegener and Ladik~\cite{Liegener:AI-87}
who use a double-$\zeta$ basis set [$(9s\,5p)\,/\,[3s\,2p]$
for fluorine and $(6s\,1p)\,/\,[2s\,1p]$ for hydrogen
to which they refer as DZP~basis set.
They observe that unit cells up to third nearest-neighbors
are sufficient to represent the Fock matrix.
My computations confirm this finding.
Liegener and Ladik~\cite{Liegener:AI-87} obtain an energy
of~$-17.57 \U{eV}$ for the top of the
valence bands and $3.57 \U{eV}$ for the bottom of the conduction
bands, \ie, their Hartree-Fock band gap amounts to~$21.14 \U{eV}$.
The top of the valence bands is only by~$0.3 \U{eV}$ smaller than
our result at the $\Gamma$~point.
However, the bottom of the conduction bands is
%
%
by~$1.78 \U{eV}$ smaller than ours.
The deviation of our result from those of Liegener
and Ladik~\cite{Liegener:AI-87} can most likely be ascribed
to the different basis sets employed, \ie, cc-pVDZ
versus DZP because the DZP basis set lacks a $d$-function
on fluorine which is present for
cc-pVDZ.~\cite{Dunning:GBS-89,Woon:GBS-93,basislib-04}
Furthermore, they use a slightly different geometry for the \HF{}~chain.

Blumen and Merkel~\cite{Blumen:EBS-77} treat the \HCl{}~chain
using a $(3s)\,/\,[1s]$~basis on hydrogen and a
$(10s\,6p)\,/\,[3s\,2p]$~basis on chlorine.
They include up to second nearest-neighbor cells to represent
the Fock matrix.
Their band structure [Fig.~1 in Ref.~\onlinecite{Blumen:EBS-77}]
agrees nicely with Fig.~\ref{fig:HCl_5cells} in this work.
We read off from the figure of Blumen and Merkel the value
$-13.4 \U{eV}$~for the top of the valence bands and
$6.9 \U{eV}$~for the bottom of the conduction bands
at the $\Gamma$~point, yielding a band
gap of~$20.3 \U{eV}$.
My corresponding values are~$-13.186 \U{eV}$
and $4.035 \U{eV}$ [Tab.~\ref{tab:cabHCl}] amounting to a
band gap of~$17.221 \U{eV}$
which is somewhat smaller than
the value of Blumen and Merkel.
Again this can most likely be ascribed to the difference
in the one-particle basis sets.

In order to investigate the effect of electron correlations
on the band structures of the chains, we utilize the CO-ADC(2,2)
method.
The convergence of several key quantities with respect
to the number of unit cells whose Wannier orbitals
are considered to form $2p1h$ and $2h1p$~configurations
in correlation calculations is summarized in
Tabs.~\ref{tab:cabHF} and \ref{tab:cabHCl} for
\HF{} and \HCl{}, respectively.
Namely, the top of the valence bands, the bottom of
the conduction bands, the band gap, and the
width of the upper and the lower valence band complexes are shown.
They are compared with the plain
Hartree-Fock data (with ``0''~unit cells for the
configuration selection).
In first place, the quasiparticle band structure of the chains is obtained
considering only a minimum fraction, a
single unit cell, in correlation calculations.%
\footnote{Due to the fact that the pseudocanonical Wannier orbitals
are delocalized over the unit cell,~\cite{Shukla:EC-96,Shukla:WF-98}
the entire cell has to be treated as
the smallest unit for configuration selection.
Hence, the application of the very fine grained selection criterion of
Refs.~\onlinecite{Buth:MC-05,Buth:AG-05}
is not sensible here.}
This causes upward shifts of the top of the valence bands
%
%
by~$1.90 \U{eV}$ for~\HF{} and
%
%
by~$0.54 \U{eV}$ for~\HCl{}, whereas the bottom of the
conduction bands shifts downwards
%
%
by~$0.22 \U{eV}$ for~\HF{} and
%
%
by~$0.60 \U{eV}$ for~\HCl{}.
Subsequently, we enlarge the configuration space to include also
configurations which extend to the nearest-neighbor cells.
This causes additional shifts of the top of the valence bands
%
%
by~$0.16 \U{eV}$ for~\HF{} and
%
%
by~$0.10 \U{eV}$ for~\HCl{}.
Moreover, shifts of the bottom of the conduction bands
%
%
of~$0.12 \U{eV}$ for~\HF{} and
%
%
of~$0.15 \U{eV}$ for~\HCl{} are observed.
Further inclusion of second and third nearest-neighbor cells
has only a minute effect.
The other quantities in Tabs.~\ref{tab:cabHF} and \ref{tab:cabHCl}
exhibit a similarly quick convergence.

The fully converged quasiparticle band structures of the
chains---configurations in up to second nearest-neighbor
cells are considered---are shown aside from the Hartree-Fock
band structures in Figs.~\ref{fig:HF_5cells} to
\ref{fig:HCl_5cells_valence}.
We observe pronounced corrections of the valence
bands of~\HF{} due to electron correlations.
Their upward shift is much larger than
the downward shift of the conduction bands.
In \HCl{} the contrary holds, however;
the value and the difference of the shifts are much smaller compared
with those of~\HF{}.
These observations match the behavior of the IPs and EAs found in the
monomers in Sec.~\ref{sec:monomer}.
Moreover, an analysis of the ground-state binding energies
of the chains
shows a similar behavior of short-range contributions of electron
correlations which are also significantly larger in~\HF{} than
in~\HCl{}.~\cite{Buth:BS-04,Buth:MC-05}
Yet the impact of the configurations from
more than a single unit cell in correlation calculations
was found to be bigger in~\HCl{}.

The observed long-range correlation corrections of the quasiparticle
band structures of the chains from more distant unit cells are,
firstly, caused by the adjustment of
the many-particle system to the extra charge of the electron which
is added to or removed from it in the one-particle Green's function.
Secondly, corrections due to van der Waals interactions
take effect.~\cite{Fulde:EC-95}

Valence and conduction bands of both compounds do not shift
symmetrically, \ie, by the same amount, upwards or downwards
[for numerical values see Tabs.~\ref{tab:cabHF} and \ref{tab:cabHCl} and the discussion of the
convergence of the CO-ADC(2,2) calculations
above].
Instead, in~\HF{}, the valence bands show a much stronger
influence due to electron correlations than the conduction bands.
The reverse trend is observed for~\HCl{}.
However, the inclusion of nearest-neighbor cells in
correlation calculations leads to essentially the same
displacement of both types of bands, \ie, the assumption
of a symmetric shift becomes
valid for long-range contributions as pointed out
in Ref.~\onlinecite{Albrecht:AA-00}.

Although the quasiparticle bands are shifted appreciably
with respect to the corresponding Hartree-Fock bands,
they essentially maintain their shape.
This observation is in contrast to the pronounced reduction of bandwidths
due to electron correlations which are
typically found in covalently bonded polymers such as
\textit{trans}-polyacetylene~\cite{Bezugly:MC-04}
and in covalently bonded crystals such as diamond
or silicon.~\cite{Grafenstein:VB-93,Grafenstein:VB-97,Albrecht:AA-00}

%
%
%
%
%
%
%
%
%
\begin{table}[tb]
  \begin{ruledtabular}
  \begin{tabular}{crrrr}
    Band No. & $\varepsilon(\Gamma )$ & $E_{2,2}(\Gamma )$ &
               $\varepsilon({\rm X})$ & $E_{2,2}({\rm X})$ \\
    \hline
           5 & -23.31 & -21.88 & -22.00 & -20.50 \\
           6 & -20.39 & -18.80 & -22.00 & -20.50 \\
           7 & -18.49 & -16.48 & -18.18 & -16.19 \\
           8 & -18.25 & -16.46 & -18.18 & -16.19 \\
           9 & -17.89 & -16.11 & -17.94 & -16.11 \\
          10 & -17.88 & -15.81 & -17.94 & -16.11 \\
          11 &   5.35 &   5.01 &   8.22 &   7.88 \\
          12 &  13.42 &  12.93 &   8.22 &   7.88
   \end{tabular}
  \end{ruledtabular}
  \caption{Energy levels of an infinite \HF{}~chain at the
           $\Gamma$ and the X~point.
           Here $\varepsilon(\vec k)$ represents the Hartree-Fock
           Bloch orbital energies and
           $E_{2,2}(\vec k)$ gives the CO-ADC(2,2) energies for
           the crystal momenta with~$\vec k = \Gamma, \rm X$
           which were determined accounting for an extension
           of the $2p1h$ and $2h1p$~configurations up to
           third nearest-neighbor cells.
           All data are given in electronvolts.}
  \label{tab:HF_band_XG}
\end{table}

To investigate the accuracy of the CO-ADC(2,2) method,
we compare it with theoretical results from many-particle perturbation
theory which has been applied
by Liegener and Ladik~\cite{Liegener:AI-87} to the
\HF{}~chain employing the DZP basis set.
For the top of the valence bands and the bottom of the
conduction bands at the $\Gamma$~point, they obtain $-15.10 \U{eV}$
and $3.05 \U{eV}$, respectively, in second order of the expansion and
$-15.09 \U{eV}$ and $3.03 \U{eV}$ in third order.
The latter
%
%
two numbers correspond to an increase of the valence
band energy by~$2.48 \U{eV}$ and a lowering of the conduction
band energy by~$0.54 \U{eV}$ with respect to the Hartree-Fock result.
In contrast, CO-ADC(2,2) yields at the $\Gamma$~point an increase
%
%
by~$2.08 \U{eV}$ and a lowering by~$0.33 \U{eV}$.
The overall agreement between CO-ADC(2,2) and
many-particle perturbation theory is satisfactory.
The deviations between the data from
both methods, as already found for the Hartree-Fock
band structure, can again be attributed
predominantly to basis set artifacts and the differences
in the geometries employed.

To the best of our knowledge no correlation calculations
of the quasiparticle band structure of \HCl{}~chains
have been performed to date.

%
%
%
%
%
%
%
%
%
\begin{table}[tb]
  \begin{ruledtabular}
  \begin{tabular}{crrrr}
    Band No. & $\varepsilon(\Gamma )$ & $E_{2,2}(\Gamma )$ &
               $\varepsilon({\rm X})$ & $E_{2,2}({\rm X})$ \\
    \hline
          13 & -17.88 & -17.14 & -17.61 & -16.88 \\
          14 & -17.24 & -16.52 & -17.61 & -16.88 \\
          15 & -13.48 & -12.85 & -13.36 & -12.73 \\
          16 & -13.39 & -12.75 & -13.36 & -12.73 \\
          17 & -13.36 & -12.72 & -13.33 & -12.69 \\
          18 & -13.19 & -12.55 & -13.33 & -12.69 \\
          19 &   4.04 &   3.28 &   5.40 &   4.56 \\
          20 &   5.78 &   4.97 &   5.40 &   4.56
   \end{tabular}
  \end{ruledtabular}
  \caption{Energy levels of an infinite \HCl{}~chain at the
           $\Gamma$ and the X~point.
           The labels are chosen as in Tab.~\ref{tab:HF_band_XG}.
           This time $2p1h$ and $2h1p$~configurations are being
           formed by the Wannier orbitals in up to second
           nearest-neighbor cells.
           All data are given in electronvolts.}
  \label{tab:HCl_band_XG}
\end{table}

\section{Conclusion}
\label{sec:conclusion}

This work deals with the quasiparticle band structures of infinite \HF{}
and \HCl{}~chains, which are determined and compared with
the energy levels of the constituting monomers.

The analysis begins with the examination of the compounds
in Hartree-Fock approximation.
The band structures of both chains turn out to be
quite similar.
However, the bands of~\HCl{} are shifted towards the
band gap with respect to the bands of \HF{}~chains.
By analyzing the ionization potentials and electron affinities of the
monomers, the origin of the individual bands of the infinite chains
is identified.
The higher outer valence bands stem from the lone
pairs on fluorine and chlorine, respectively, and the lower
outer valence bands originate from the
hydrogen bonding.
The latter bands also exhibit less dispersion in~\HCl{}
compared with~\HF{} due to the weaker bonding.

Electron correlation calculations are performed to
determine accurate energy levels of the HF and HCl~molecules
and precise quasiparticle band structures
of the \HF{} and \HCl{} chains.
To this end, we use the \emph{ab initio} Green's function theory
crystal orbital algebraic diagrammatic construction for
Wannier orbitals.
The strict second order approximation of the off-diagonal Fock
matrix elements and the two-electron contributions is applied,
which is dubbed CO-ADC(2,2)~method.~\cite{Buth:MC-05,Buth:AG-05}
Setting out from these equations, we make the transition
to the corresponding method for
molecules, termed~ADC(2,2), by eliminating the
lattice summations.
The configuration space of the correlation calculations
is found to converge rapidly;
only nearest-neighbor cells need to be regarded
to determine the band gap of the chains
%
%
%
%
within~$0.008 \U{eV}$ of the value obtained in
the largest calculations performed.
To assess the accuracy of the ADC(2,2) ionization potentials
of the monomers, we compare with
ADC(3)~data;
a reassuring agreement is found.
Furthermore, the quasiparticle energies for the \HF{}~chain
compare satisfactorily with data from many-particle perturbation
theory,~\cite{Liegener:AI-87}
thus corroborating the accuracy of the CO-ADC(2,2)~scheme.

The inclusion of correlation effects leads to a
pronounced shift towards the band gap of the quasiparticle bands
with respect to the Hartree-Fock bands.
The shift of the valence bands is noticeably larger in the
\HF{}~chain compared with the \HCl{}~chain;
the reverse trend is observed for the conduction bands.
These observations match the behavior of the energy levels
of the monomers.
In contrast to many other compounds, \eg, covalently
bound polymers~\cite{Bezugly:MC-04} or
crystals,~\cite{Grafenstein:VB-93,Grafenstein:VB-97,Albrecht:AA-00}
the bandwidth of the chains does not decrease substantially due to
correlation effects but increases slightly.

\begin{table}[tb]
  \begin{ruledtabular}
  \begin{tabular}{rccccc}
    Cells & $E_{\rm top,v}$ & $E_{\rm bottom,c}$ & $E_{\rm gap}$ &
    $\Delta E^{\,<}_{{\rm F}\,2p}$ & $\Delta E^{\,>}_{{\rm F}\,2p}$ \\
    \hline
    0 & -17.876 & 5.347 & 23.223 & 2.924 & 0.610 \\
    1 & -15.975 & 5.129 & 21.104 & 2.993 & 0.654 \\
    3 & -15.815 & 5.011 & 20.825 & 3.078 & 0.664 \\
    5 & -15.813 & 5.006 & 20.819 & 3.078 & 0.664 \\
    7 & -15.812 & 5.005 & 20.817 & 3.078 & 0.664
  \end{tabular}
  \end{ruledtabular}
  \caption{Convergence of the fundamental band
           gap~$E_{\rm gap}$ and the bandwidth of the lower
           and the upper F$\,2p$~valence band
           complexes, $\Delta E^{\,<}_{{\rm F}\,2p}$ and
           $\Delta E^{\,>}_{{\rm F}\,2p}$,
           respectively, of a \HF{}~chain with
           respect to the number of unit cells used to determine
           the quasiparticle band structures.
           ``Cells''~designates the number of
           unit cells taken into account in
           CO-ADC(2,2) calculations, where zero
           refers to the original Hartree-Fock results.
           Unity denotes the $2p1h$ and $2h1p$~configurations
           from the origin cell only in correlation calculations.
           Three, five, and seven indicate the additional inclusion of
           nearest-, second nearest- and third nearest-neighbor cells,
           respectively. The top of the valence
           bands~$E_{\rm top,v}$ and the bottom of
           the conduction bands~$E_{\rm bottom,c}$
           are both situated at the $\Gamma$~point.
           All data are given in electronvolts.}
  \label{tab:cabHF}
\end{table}

Building on top of this work, one should investigate
carefully the lower lying, inner valence regime
of the chains where intriguing many-particle effects are to be
expected.~\cite{Santra:ED-01,Cederbaum:CE-86,Buth:IO-03,Ohrwal:FS-04}
%
%
%
%
%
%
%
%
%
%
%
%
%
%
%
%
%
%
%
%
%
%
%
%
%
%
%
%
%
%
%
%
%
%
\begin{table}[th]
  \begin{ruledtabular}
  \begin{tabular}{rccccc}
    Cells & $E_{\rm top,v}$ & $E_{\rm bottom,c}$ & $E_{\rm gap}$ &
    $\Delta E^{\,<}_{{\rm Cl}\,3p}$ & $\Delta E^{\,>}_{{\rm Cl}\,3p}$ \\
    \hline
    0 & -13.186 & 4.035 & 17.221 & 0.641 & 0.294 \\
    1 & -12.651 & 3.434 & 16.085 & 0.620 & 0.288 \\
    3 & -12.554 & 3.280 & 15.834 & 0.627 & 0.299 \\
    5 & -12.550 & 3.277 & 15.827 & 0.627 & 0.299
  \end{tabular}
  \end{ruledtabular}
  \caption{Convergence of the fundamental band
           gap~$E_{\rm gap}$ and the bandwidth of the lower
           and upper Cl$\,3p$~valence band
           complexes, $\Delta E^{\,<}_{{\rm Cl}\,3p}$ and
           $\Delta E^{\,>}_{{\rm Cl}\,3p}$,
           respectively, of a \HCl{}~chain with
           respect to the number of unit cells used to determine
           the quasiparticle band structures.
           The symbols are chosen as in Tab.~\ref{tab:cabHF}.
           All data are given in electronvolts.}
  \label{tab:cabHCl}
\end{table}
This is indicated by the studies of oligomers by
Deleuze~\etal,~\cite{Deleuze:SC-92,Deleuze:CE-96,%
Deleuze:SB-96,Deleuze:PB-96,Deleuze:VO-99}
which model the corresponding infinite periodic chains.
They serve to investigate the evolution of strong electron correlations
with an increasing length of the oligomers.
Moreover, intermolecular electronic decay processes, \eg,
Ref.~\onlinecite{Buth:IM-03} (and references therein)
found recently in atomic and molecular clusters, including clusters
of HF~molecules,~\cite{Zobeley:HE-98} present challenging perspectives.
Further emerging effects, when making the transition from
the monomer of an infinite chain, have been discovered in
the model oligomer studies of Refs.~\onlinecite{Deleuze:AG-91,Deleuze:SC-92,Ortiz:EB-94,Deleuze:CE-96,%
Deleuze:SB-96,Deleuze:PB-96,Deleuze:VO-99,Golod:VC-99,Deleuze:TN-99}
which deserve a close look within an infinite chain treatment.
Particularly, the examination of these effects in three-dimensional
crystals offers a fascinating perspective but also represents a great challenge.

\begin{acknowledgments}
I am indebted to Thomas Sommerfeld for providing a
complex-symmetric band-Lanczos code and helpful advice.
Francesco Tarantelli kindly granted access to his
molecular ADC~program.
Moreover, I would like to thank Martin Albrecht for fruitful
discussions.
This work was partly supported by a Feodor Lynen Research Fellowship
from the Alexander von Humboldt Foundation.
\end{acknowledgments}

\end{document}